\documentclass[aps,onecolumn,12pt,superscriptaddress,letterpaper,amsmath,amssymb,floatfix,pra]{revtex4-2}

\usepackage{graphicx}
\usepackage{dcolumn}
\usepackage{bm}
\usepackage{subcaption}
\usepackage{hyperref}
\def\theYear{\the\year}

\begin{document}
\setcounter{page}{1}

\title{Compton Edge Convolutional Model and Algorithm for Energy-channel Calibration}

\author{Yanbao Zhang}
\email{zhangyanbiao@stu.scu.edu.cn}
\affiliation{Nuclear Engineering, College of Physics, Sichuan University, China}
\author{Fanjie Zeng}
\affiliation{Nuclear Engineering, College of Physics, Sichuan University, China}
\author{Dehua Kong}
\affiliation{Center for Theoretical Physics, College of Physics, Sichuan University, China}
\author{Lian Lei}
\affiliation{Center for Theoretical Physics, College of Physics, Sichuan University, China}
\author{Zhonghai Wang}
\email{zhonghaiwang@scu.edu.cn}
\affiliation{Nuclear Engineering, College of Physics, Sichuan University, China}

\date{\today}

\begin{abstract}
Scintillation detectors are essential tools for radiation measurement, but calibrating them accurately can be challenging, especially when full-energy peaks are not prominent. This is common in detectors like plastic scintillators. Current methods for calibrating these detectors often require manual adjustments.
To address this, we propose a new method called the convolution model. This model accurately calibrates the energy-channel relationship of the Compton edge in various detectors. We tested it with plastic scintillator BC408, NaI crystal, and LaBr$_3$ crystal.
Using ${}^{137}$Cs radioactive sources, we calibrated NaI and LaBr$_3$ detectors using full-energy peaks, then applied the convolution model to fit the Compton edge. Our results show errors within 1\% when compared to full-energy peak calibration.

\end{abstract}
\maketitle
\vspace{-0.8cm}
\section{Introduction}

Scintillation detectors play a pivotal role in various fields, including radiation dosimetry, nuclear physics, environmental monitoring, and medical imaging \cite{Wang_2023}. These detectors convert incident radiation into flashes of light, which are then detected and converted into electrical signals for analysis \cite{PhysRev.21.483}. The accuracy and reliability of scintillation detectors depend largely on the calibration of their energy-channel relationship, which correlates the detected signals with the energy of the incident radiation. \cite{PhysRev.22.409}

The conventional method for calibrating scintillation detectors involves using full-energy peaks, which represent the complete absorption of incident radiation within the detector material. However, certain types of scintillation detectors, such as plastic scintillators, may exhibit pulse height spectra dominated by Compton plateaus and edges, with full-energy peaks being either weak or absent. This poses a significant challenge for calibration, as traditional methods reliant on full-energy peaks may not be applicable.

Calibrating the energy-channel relationship of scintillation detectors with predominantly Compton spectra typically involves selecting points along the Compton descent, usually between 50\% to 70\% of the maximum amplitude, for calibration \cite{article2}. However, this approach often requires manual adjustment and may lack universality across different detector types, leading to inaccuracies and inconsistencies. \cite{Deng_2022}

To address this challenge, this paper proposes a novel convolution model for accurately fitting the energy-channel relationship of the Compton edge in scintillation detectors. Unlike traditional methods, the convolution model offers a systematic and adaptable approach that does not rely on the presence of full-energy peaks. \cite{JEON20198}. By leveraging the unique characteristics of Compton plateaus and edges, the model aims to provide a universal calibration framework applicable to a wide range of scintillation detectors \cite{Wang_2023}.

In this study, experiments were conducted using three types of scintillation detectors: plastic scintillator BC408, NaI crystal, and LaBr$_3$ crystal, representing a diverse range of detector materials. The proposed convolution model was applied to fit the energy-channel relationship of the Compton edge for each detector type. Experimental validation was performed by comparing the resulting energy-channel functions with those obtained from full-energy peaks calibration, with errors within 1\% considered acceptable.

The development of accurate and universally applicable calibration methods for scintillation detectors is crucial for ensuring the reliability and precision of radiation detection and measurement \cite{article3}. The proposed convolution model represents a significant advancement in this endeavor, offering a practical solution for calibrating scintillation detectors with predominantly Compton spectra. By enhancing the accuracy and versatility of calibration techniques, this model has the potential to broaden the scope of applications for scintillation detectors and facilitate advancements in radiation detection technology \cite{Abbas_2006}.

The Compton effect is a phenomenon where the wavelength of a scattered photon varies with the scattering angle. During this effect, a photon collides elastically with an electron outside the nucleus of an atom. As a result, part of the energy is transferred to the electron, causing it to break away from the atom and become a recoil electron. This changes the energy and direction of motion of the scattered photon \cite{article3}. 
For energy detection of $\gamma$ rays, if the first time a $\gamma$ photon occurs in the detection region is the Compton effect and the scattered photon escapes from the detection region, the recoil electron will form a continuous spectrum with energy from $0 \sim E_\gamma /\left(1+(1 / 4) E_\gamma\right)$, i.e. Compton plateau \cite{JEON20198}. 

\section{Methods}
Scintillation Detectors and Experimental Setup
Three types of scintillation detectors were utilized in this study: plastic scintillator BC408, NaI crystal, and LaBr$_3$ crystal. These detectors were chosen to represent a diverse range of scintillation materials commonly used in radiation detection applications. Each detector was coupled to appropriate photomultiplier tubes (PMTs) to convert scintillation light into electrical signals.

The experimental setup consisted of a source holder housing ${}^{137}$Cs radioactive sources emitting gamma rays at 662 keV. The detectors were positioned at a fixed distance from the source holder to ensure consistent irradiation conditions throughout the experiments.

\subsubsection{Theoretical Image of Compton Recoil Electron Angular Distribution}

\begin{figure}[htbp]
    \centering
    \includegraphics[width=0.5\linewidth]{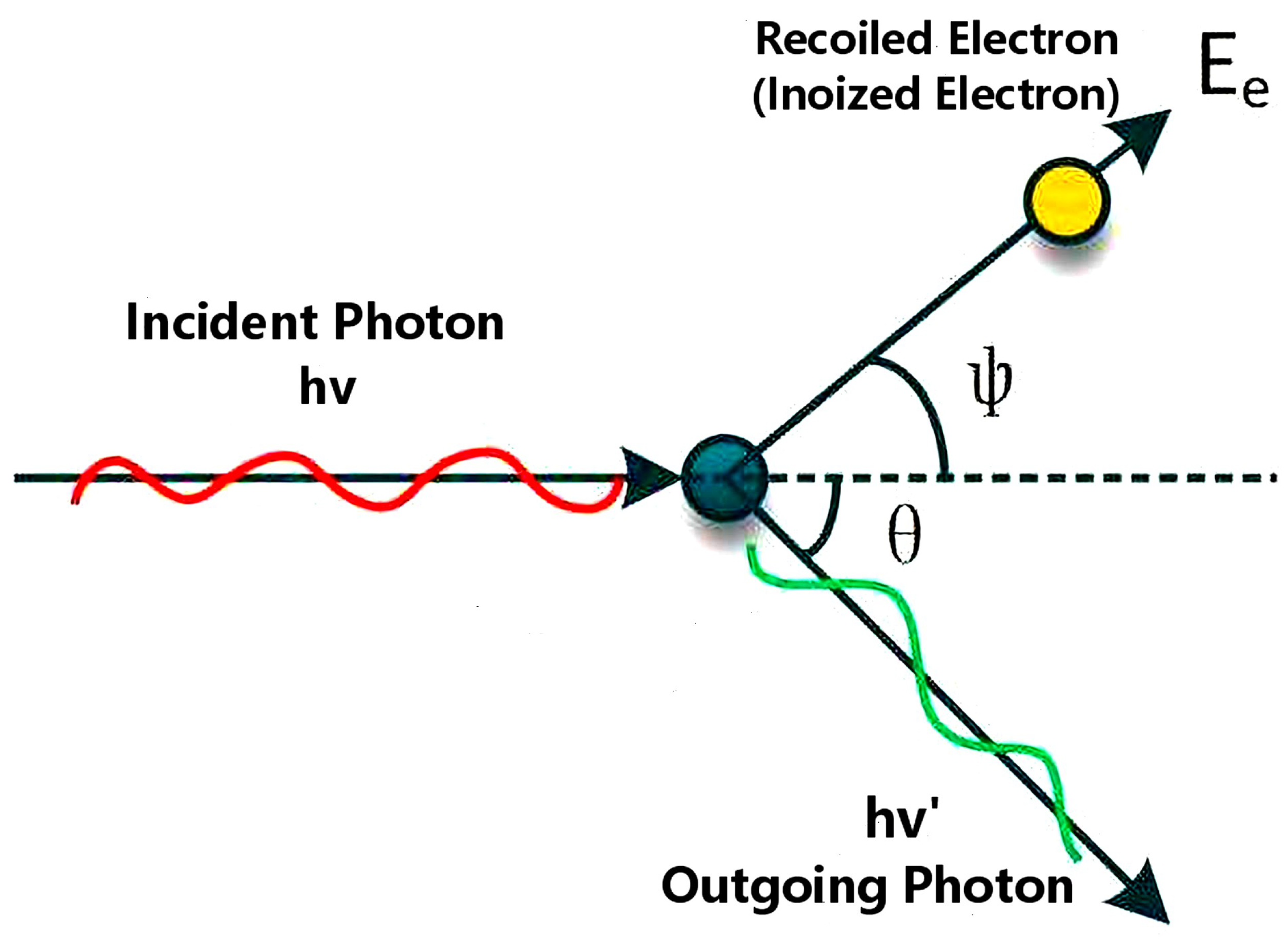}
    \caption{Schematic diagram of the Compton effect}
    \label{fig:Compton}
\end{figure}

In the Compton scattering process, $\mathrm{h} v$ denotes the energy of the incident photon, $\mathrm{h} v^{\prime}$ denotes the energy of the scattered photon, and the energy of the recoil electrons is denoted as $E_{\mathrm{e}}$ as shown in \autoref{fig:Compton}. It is easy to derive the relationship between the scattered photon energy $\mathrm{h} v^{\prime}$ and the scattering angle $\theta$ as follows.
\begin{equation}
    \mathrm{h} v^{\prime}=\frac{\mathrm{h} v}{1+\alpha(1-\cos \theta)}
\end{equation}
where $\alpha=\mathrm{h} v /\left(\mathrm{m}_{\mathrm{e}} \mathrm{c}^2\right)$, is the ratio of the energy of the incident $\gamma$ photon to the rest mass of the electron. According to equation (1), the Compton electron has a maximum kinetic energy under the condition $\theta=\pi$, which is.
\begin{equation}
   E_{\mathrm{e}(\max )}=\frac{2 \alpha}{1+2 \alpha} \mathrm{h} v_0
\end{equation}

During Compton scattering, the kinetic energy of the recoil electrons varies from 0 to $E_{\mathrm{e}(\max )}$, and $E_{\mathrm{e}(\max )}$ determines the right edge of the Compton plateau in the $\gamma$-ray spectrum. Of course, different kinetic energies of recoil electrons correspond to different recoil angles, and there is a probability distribution of recoil electrons falling at different recoil angles, which is the angular distribution of Compton recoil electrons.

\begin{figure}
    \centering
    \includegraphics[width=0.6\textwidth]{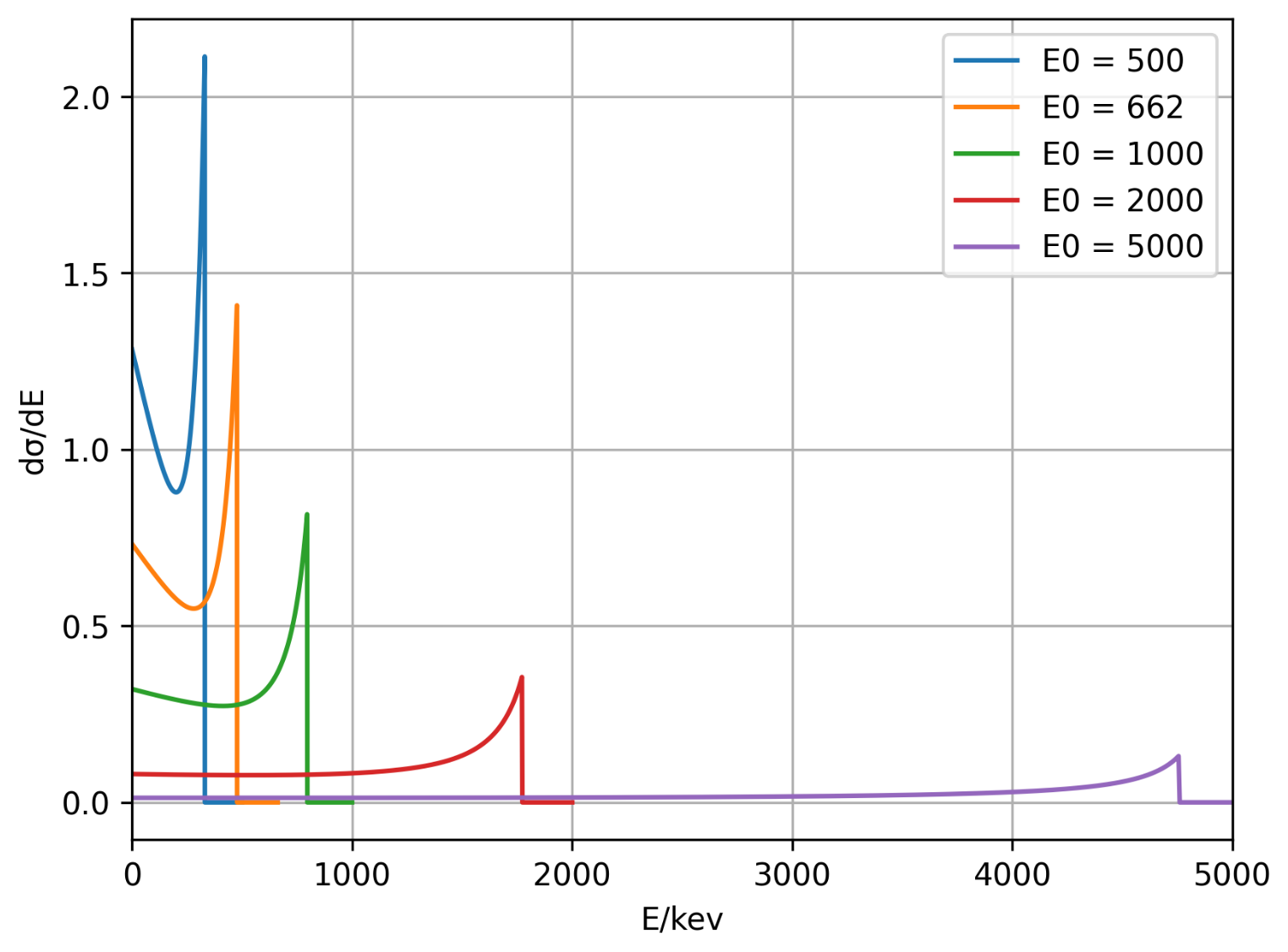}
    \caption{Differential Cross Section for Recoil Electron Energy}
    \label{fig:Cross-Section}
\end{figure}

\begin{figure}[htbp]
    \centering
    \begin{subfigure}{0.49\textwidth}
        \includegraphics[width=\textwidth]{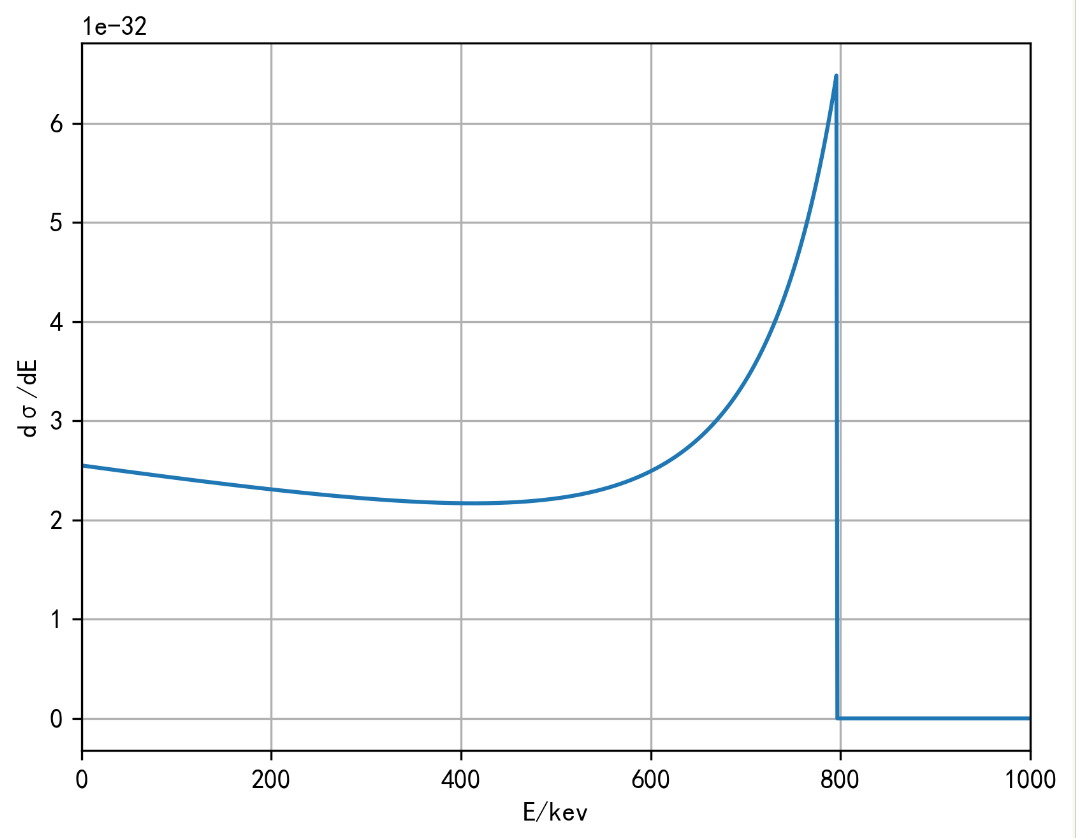}
        \caption{Recoil electron differential cross section}
    \end{subfigure}
    \begin{subfigure}{0.49\textwidth}
        \includegraphics[width=\textwidth]{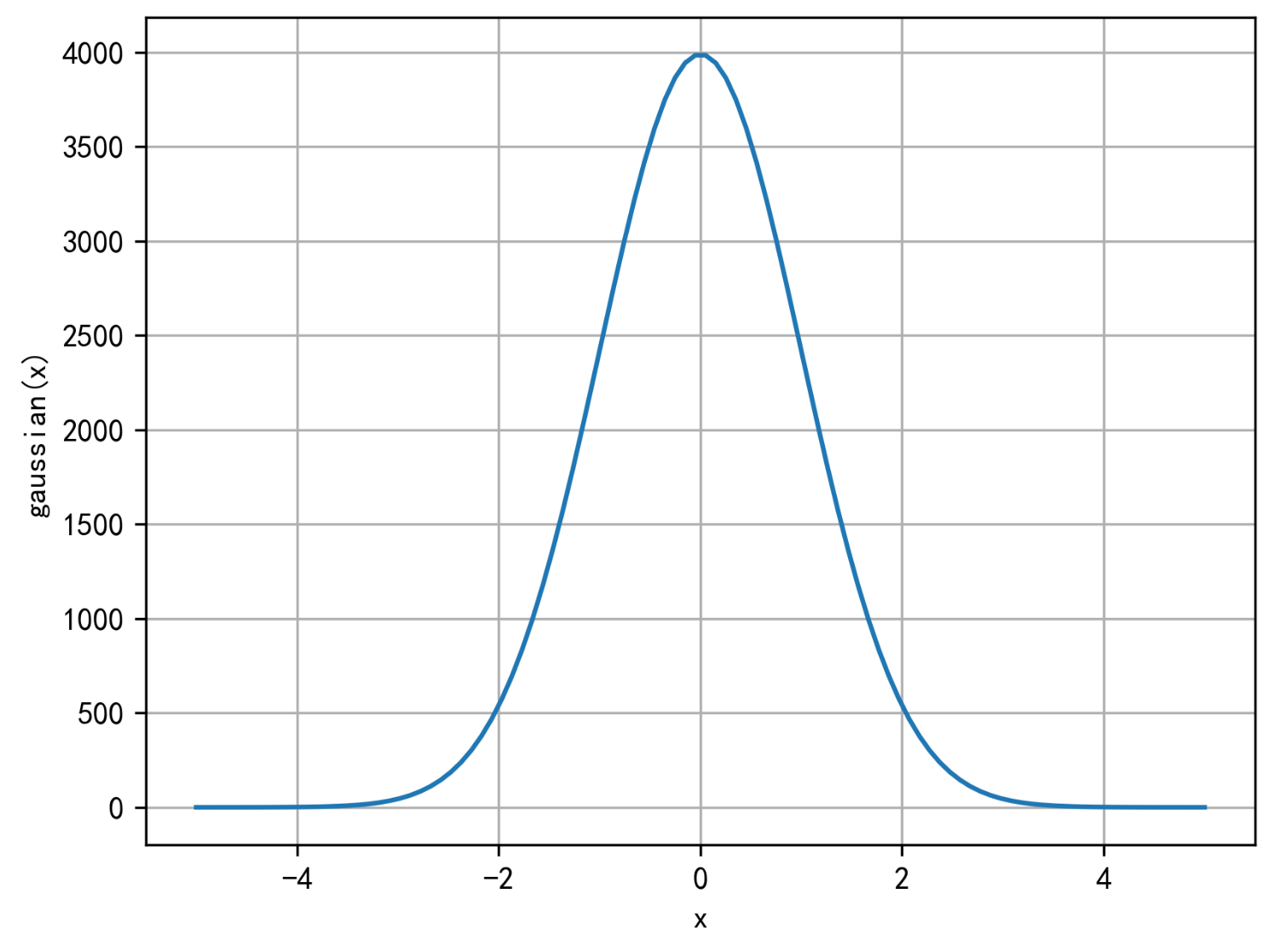}
        \caption{Gaussian convolution kernel}
    \end{subfigure}
    \caption{convolution process}
    \label{fig:conv1}
\end{figure}

The theoretical
image of the angular distribution of Compton 
recoil electrons can be derived from the differential cross 
section of Compton scattered photons given by the Compton scattering theory in quantum electrodynamics, i.e., it is based on the well-known Klein-Nishina formula \cite{article}.
\begin{equation}
    \frac{\mathrm{d} \sigma}{\mathrm{d} \Omega_\theta}=\mathrm{r}_0^2\left(\frac{1}{1+\alpha(1-\cos \theta)}\right)^2\left(\frac{1+\cos ^2 \theta}{2}\right)\left(1+\frac{\alpha^2(1-\cos \theta)^2}{\left(1+\cos ^2 \theta\right)[1+\alpha(1-\cos \theta)]}\right),
\end{equation}

where $\mathrm{r}_0=2.818 \times 10^{-15} \mathrm{~m}$, is the classical electron radius, $\alpha=\mathrm{h} v /\left(\mathrm{m}_{\mathrm{e}} \mathrm{c}^2\right)$, 
Compton The differential cross section of a scattered photon is denoted by $\mathrm{d} \sigma / \mathrm{d} \Omega_\theta$, which is expressed in $\mathrm{cm}^2 \cdot \mathrm{sr}^{-1}$, and $\mathrm{d} \sigma / \mathrm{d} \Omega_\theta$. 
The physical meaning of $Omega_\theta$ is that for a medium containing only one electron, the probability of a scattered photon scattering into a steradian angle of unit radian in the direction of $\theta$ when a photon is incident perpendicularly to the unit area. This curve is shown in \autoref{fig:Cross-Section} and \autoref{fig:conv1}.

Scattered photons at a certain scattering angle correspond to recoil electrons at a certain recoil angle. Scattered photons falling into $\theta \sim \theta+\mathrm{d} \theta$ and recoil electrons falling into $\varphi \sim \varphi+\mathrm{d} \varphi$ are the same random event, and the probability of both is the same. Therefore, there is the following relationship.

\begin{equation}
    \frac{\mathrm{d} \sigma}{\mathrm{d} \Omega_{\mathrm{e}}} 2 \pi \sin \varphi \mathrm{d} \varphi=\frac{\mathrm{d} \sigma}{\mathrm{d} \Omega_\theta} 2 \pi \sin \theta \mathrm{d} \theta,
\end{equation}

where $\mathrm{d} \sigma / \mathrm{d} \Omega_{\mathrm{e}}$ is the differential cross-section of a recoil electron, which is the probability of a recoil electron falling within a steradian angle of unit radian in the direction of $\varphi$, which is also in units of $\mathrm{cm}^2 \cdot \ mathrm{sr}^{-1}$. Combining Eq. (3) with Eq. (8) below yields the differential cross-section expression for the Compton recoil electron as
\begin{equation}
    \frac{\mathrm{d} \sigma_{e, e}}{\mathrm{~d} E_e}=\frac{\pi r_0^2}{\alpha^2 m_0 c^2} \cdot\left[2+\left(\frac{E_e}{h \nu-E_e}\right)^2\left(\frac{1}{\alpha^2}-\frac{2\left(h \nu-E_e\right)}{\alpha E_e}+\frac{h_\nu-E_e}{h_\nu}\right)\right]
\end{equation}
We use $\mathrm{d} \sigma / \mathrm{d} \varphi$ to denote the distribution of recoil electrons with respect to the recoil angle, which refers to the probability that the recoil electrons fall within the unit recoil angle in the direction of $\varphi$. Using equation (5), the expression for $\mathrm{d} \sigma / \mathrm{d} \varphi$ can be calculated as follows.

The energy-channel calibration for NaI and LaBr$_3$ crystals was performed using full-energy peaks. A known gamma-ray source emitting energies at characteristic peaks was used to irradiate the detectors. The pulse height spectra were acquired using data acquisition systems, capturing the response of the detectors to gamma-ray interactions. The centroids of the full-energy peaks in the pulse height spectra were identified and correlated with the corresponding gamma-ray energies.
A linear regression model was then employed to establish the energy-channel relationship, mapping channel numbers to gamma-ray energies for each detector. This calibration procedure provided the baseline energy-channel functions for NaI and LaBr$_3$ crystals.

\subsection{Convolution model}
In high-energy physics measurements, the finite resolution of the measurement system always causes a broadening of the energy spectrum. In order to eliminate this effect and obtain the true spectrum (ideal spectrum), the measured spectrum is often deconvolved according to the resolution function of the instrument \cite{Paatero1974-PAADIC}.
The functions of common physical devices, such as amplifiers, filters, optical instruments, etc., can be summarised as follows.
\begin{equation}
    \text { input } f(t) \rightarrow \text { device } \rightarrow \text { output } h(t)
\end{equation}
\begin{figure}[h]
    \centering
    \includegraphics[width=0.6\textwidth]{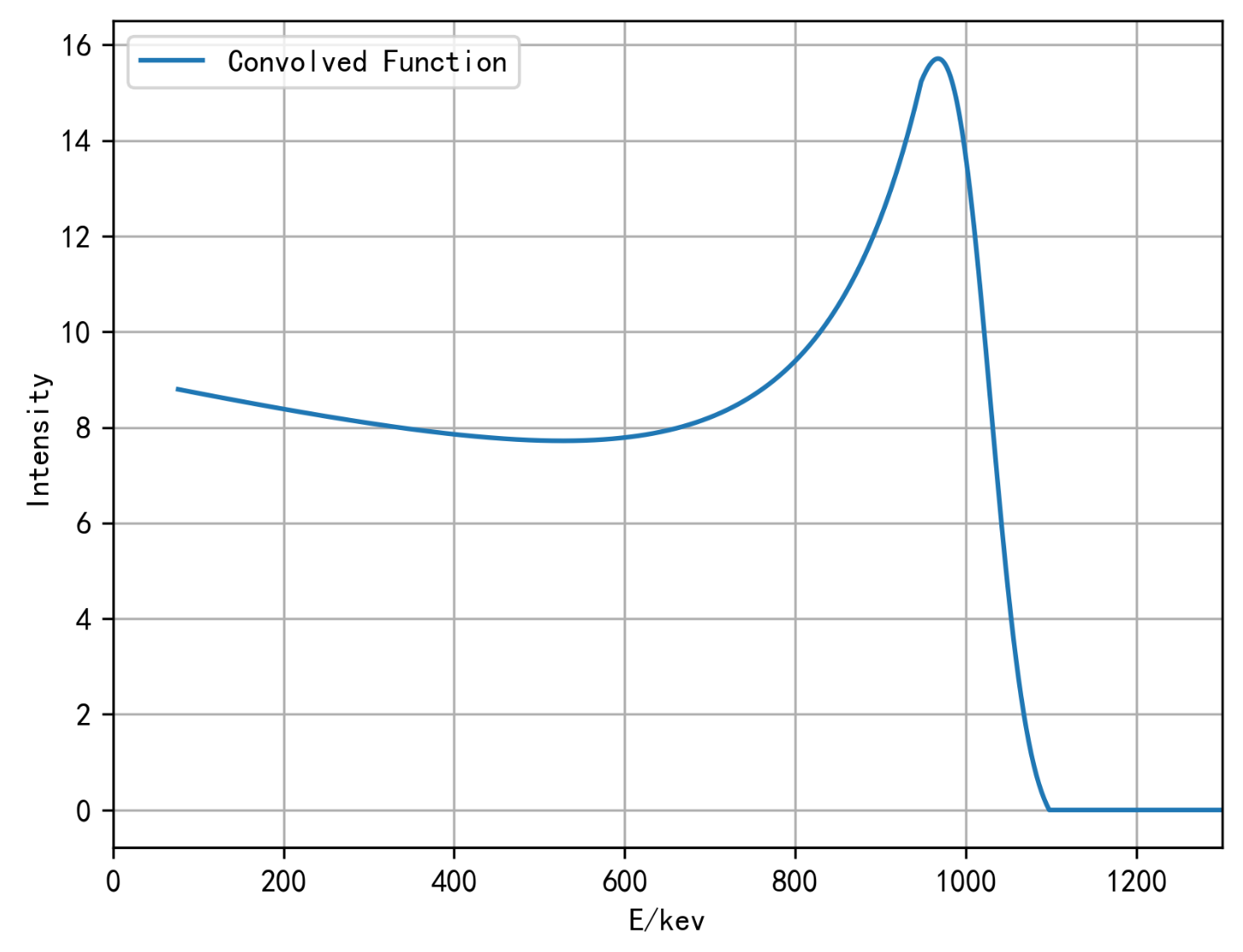}
    \caption{Differential Cross Section for Recoil Electron Energy}
    \label{fig:conv2}
\end{figure}
The output $h(t)$ depends not only on the input $f(t)$, but also on the performance of the device itself. In general. The device has the following two properties \cite{Philip1983DeconvolutionOG}.
1) Linearity
If $f_1(t), f_2(t) \rightarrow h_1(t), h_2(t)$, then $\alpha f_1(t)+\alpha_2 f_2(t) \rightarrow \alpha_1 h_1(t)+\alpha_2 h_2(t)$, which means that the output signals have the same linear iterative pattern as the input signals. This means that the output signal has the same linear iteration pattern as the input signal.
2) Translation invariance
If $f(t) \rightarrow h(t)$, then $f(t-\tau) \rightarrow h(t-\tau)$.
Let the input be $\delta(t)$, and the output be $g(t)$. The $g(t)$ is called the resolution function (characteristic function) of the device. It directly reflects the resolution of the device. Obviously. If the input is $\delta(t-\tau)$, then the output is $g(t-\tau)$.
By the nature of the $\delta$ function, any input $f(t)$ can always be written as
\begin{equation}
    f(t)=\int_{-\infty}^{+\infty} f(\tau) \delta(t-\tau) \mathrm{d} \tau, \quad-\infty<t<+\infty, \quad-\infty<t<+\infty,
\end{equation}
And this process is shown in \autoref{fig:conv1}. The corresponding output is
\begin{equation}
   h(t)=\int_{\infty}^{+\infty} f(\tau) g(t-\tau) \mathrm{d} \tau \equiv f(t) * g(t) .
 \end{equation}

This means. The output function $h(t)$ is the convolution of the input function $f(t)$ with the device resolution function $g(t)$. For the discrete spectrum, equation (1) reads
\begin{equation}
    h(i)=\sum_{i=-\infty}^{+\infty} f(j) g(i-j) .
\end{equation}

Compton scattering can be regarded as a collision of a photon with energy $h \nu$ and a stationary free electron with mass $m_0$, according to the relativistic law of conservation of energy and momentum, there are
\begin{equation}
\begin{aligned}
& \mathrm{h} \nu+E_0=\mathrm{h}_{\nu^{\prime}}{ }^{\prime}+E, \\\\
& p_\lambda^2+p_{\lambda^{\prime}}^2-2 p_\lambda p_{\lambda^{\prime}} \cos \theta=p^2,
\end{aligned}
\end{equation}
The parameters are shown in \autoref{tab:tab1}
\begin{figure}[htbp]
    \centering
    \begin{subfigure}{0.32\textwidth}
        \includegraphics[width=\textwidth]{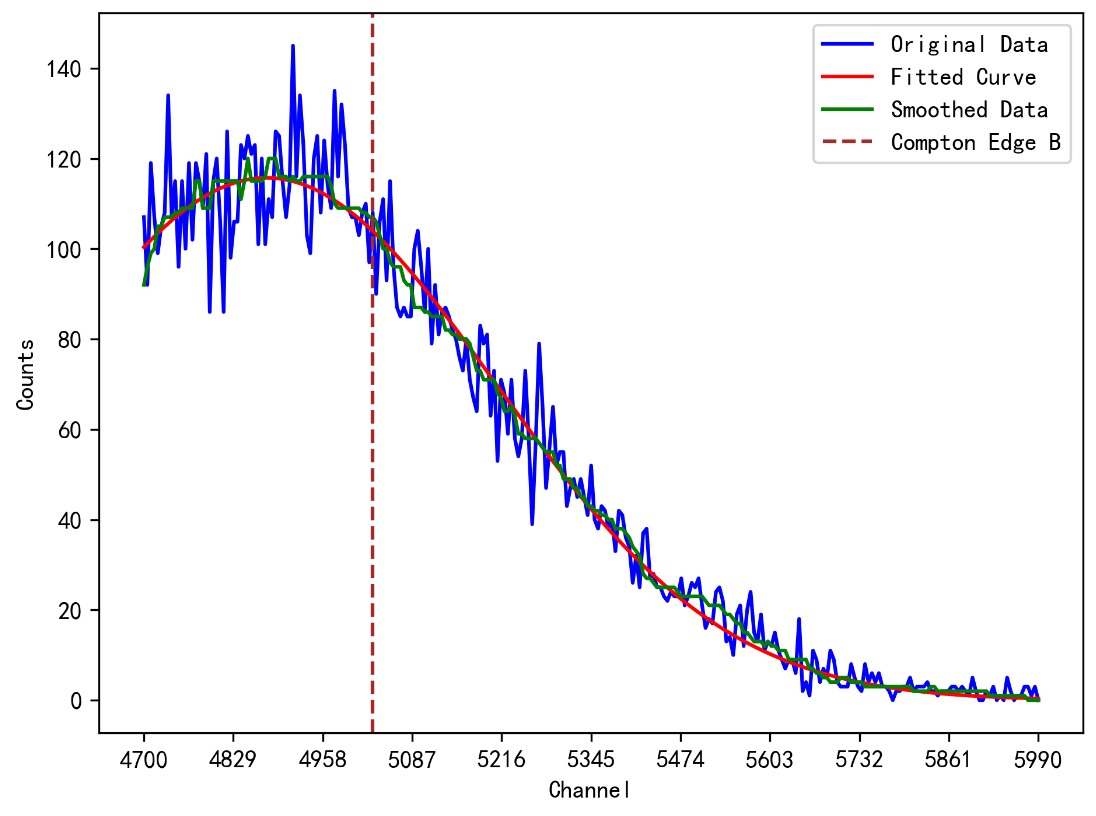}
        \caption{BC408 and ${}^{22}$Na Fitting}
    \end{subfigure}
    \begin{subfigure}{0.32\textwidth}
        \includegraphics[width=\textwidth]{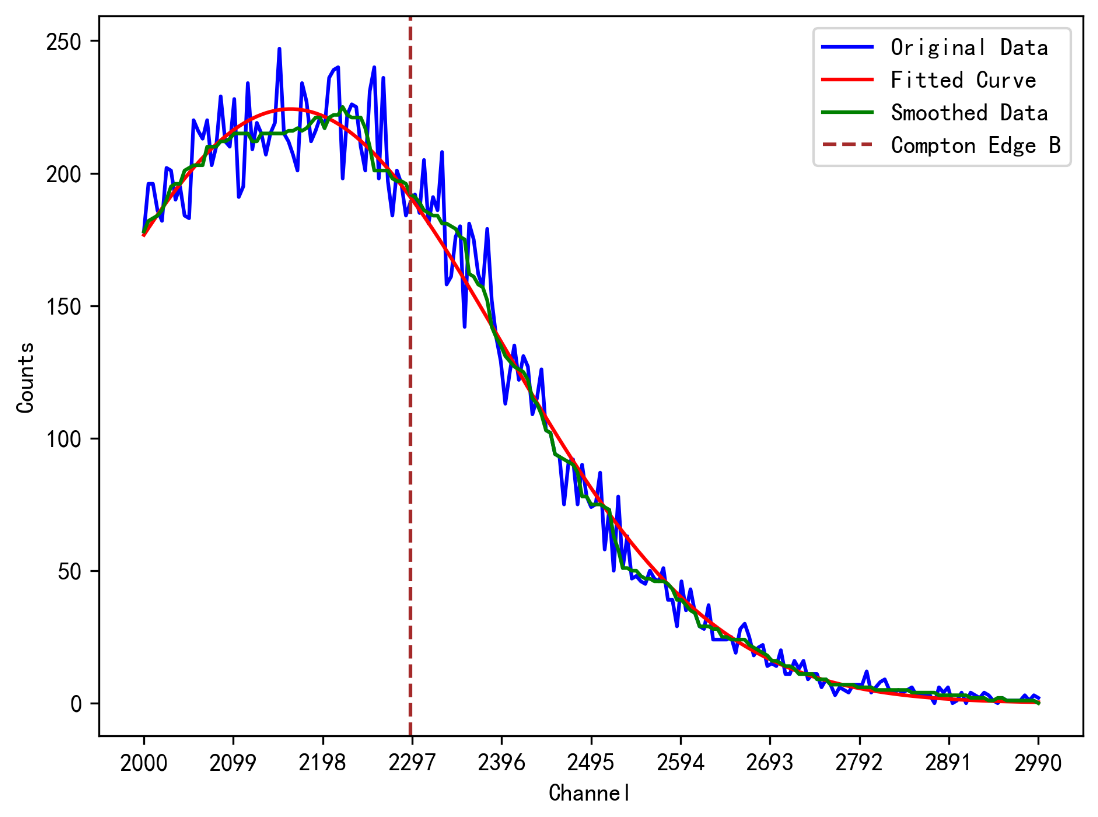}
        \caption{BC408 and ${}^{137}$Cs Fitting}
    \end{subfigure}
    \begin{subfigure}{0.32\textwidth}
        \includegraphics[width=\textwidth]{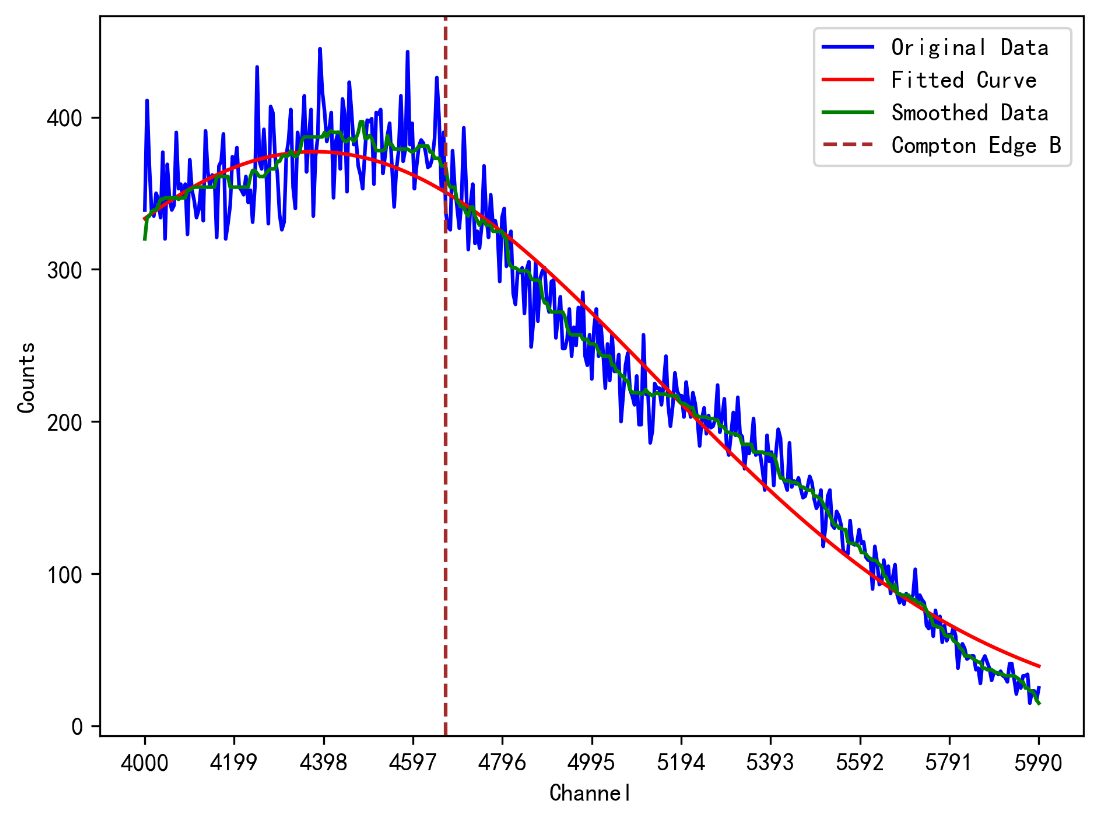}
        \caption{BC408 and ${}^{60}$Co Fitting}
    \end{subfigure}

    \caption{Compton Edge Fitting}
    \label{fig:fit1}
\end{figure}

where: $E$ and $p$ are the energy and momentum of the recoil electron respectively; $E_0=m_0 c^2$ is the rest energy of the electron; and $p_\lambda$ and $p_{\lambda^{\prime}}$ are the momentum before and after the collision of the photon respectively. This gives the energy of the recoil electron as
\begin{equation}
    E=\frac{\mathrm{h}_\nu}{1+E_0 /(\mathrm{h} \nu(1-\cos \theta))} 
\end{equation}

From the union of Eq. (1) and Eq. (4), we have $\frac{\mathrm{d} \sigma_c}{\mathrm{~d} E}=\frac{\mathrm{d} \sigma_{\mathrm{c}}}{\mathrm{d} \Omega} \frac{\mathrm{d} \Omega}{\mathrm{d} \theta} \frac{\mathrm{d} \theta}{\mathrm{d} E}$, i.e., the recoil electron energy differential cross-section, is shown in the theoretical curve in \autoref{fig:conv2}.

The image plotted when $E_\gamma$ = 1000 kev is as follows:

We define the above function as f(E)
\begin{equation}
    \left[g(E)=\frac{A}{\sigma \sqrt{2 \pi}} e^{-\frac{(E-\mu)^2}{2 \sigma^2}}\right]
\end{equation}

Let $f(E)$ and $g(E)$ be convolved to obtain the function $h(E)$.

\begin{equation}
    h(E)=(f * g)(E)=\int_{-\infty}^{\infty} f\left(E^{\prime}\right) g\left(E-E^{\prime}\right) d E^{\prime}
\end{equation}
The convolution function image obtained when A=10000, $\sigma$=1,E0=1000kev is shown in \autoref{fig:conv1}(a).

Data analysis was conducted using appropriate software tools for signal processing, spectrum analysis, and statistical calculations. The energy-channel functions derived from both calibration methods were compared using statistical metrics to assess the accuracy and consistency of the convolution model. The fitting results are shown in \autoref{fig:fit1} and \autoref{fig:fit2}.

\section{Conclusion}

In this study, we have proposed a novel convolution model for accurately fitting the energy-channel relationship of the Compton edge in scintillation detectors, particularly in cases where full-energy peaks are not prominently observed. The convolution model offers a systematic and adaptable approach that bypasses the reliance on full-energy peaks, providing a universal calibration framework applicable to a wide range of detector types.

Experimental validation using plastic scintillator BC408, NaI crystal, and LaBr$_3$ crystal detectors demonstrated the effectiveness and versatility of the proposed model. By comparing the resulting energy-channel functions with those obtained from full-energy peaks calibration, errors within 1\% were observed, affirming the accuracy and reliability of the convolution model.
\begin{figure}[htbp]
    \centering
    \begin{subfigure}{0.48\textwidth}
        \includegraphics[width=\textwidth]{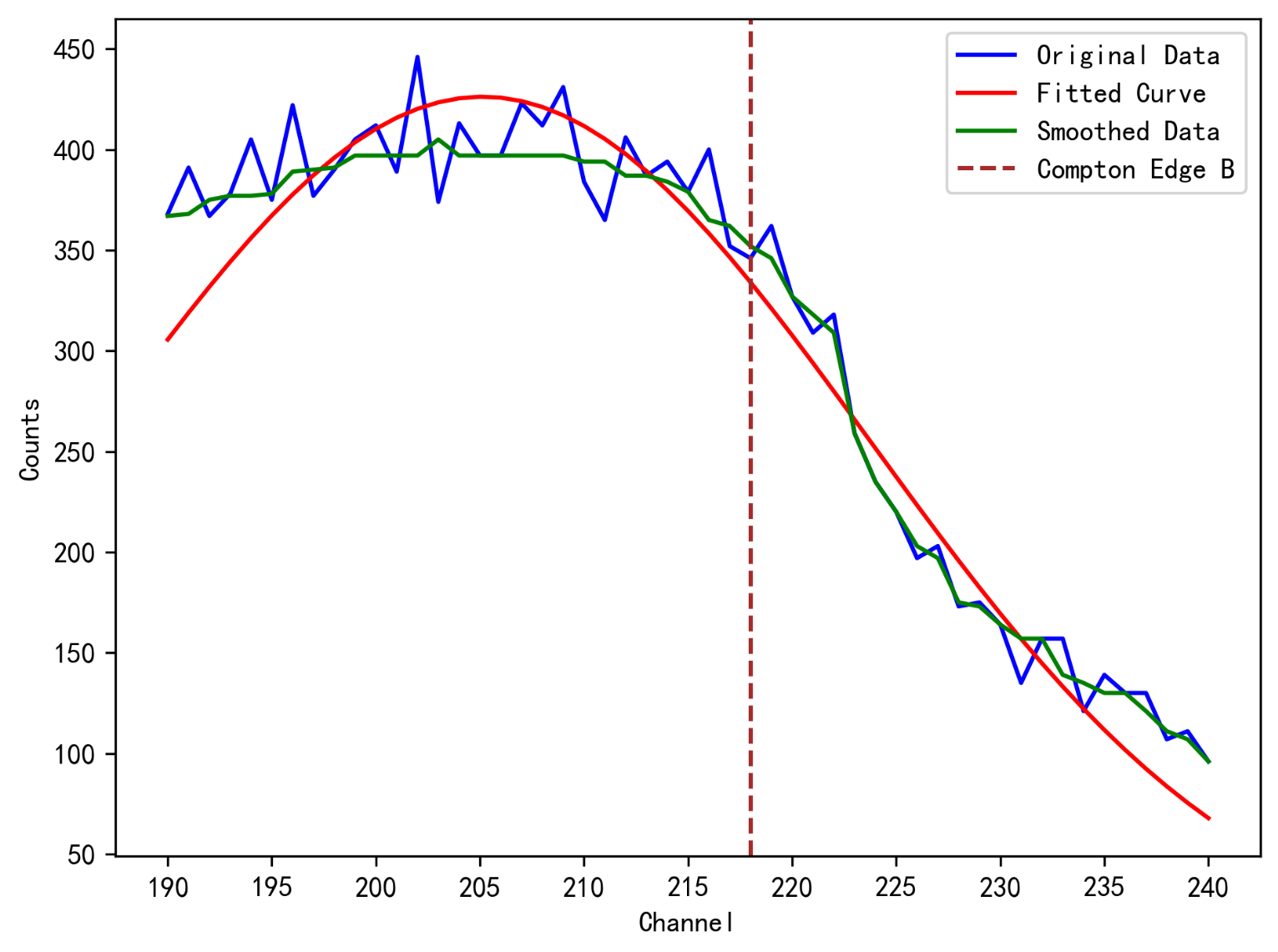}
        \caption{LaBr$_3$ and ${}^{137}$Cs Fitting}
    \end{subfigure}
    \begin{subfigure}{0.48\textwidth}
        \includegraphics[width=\textwidth]{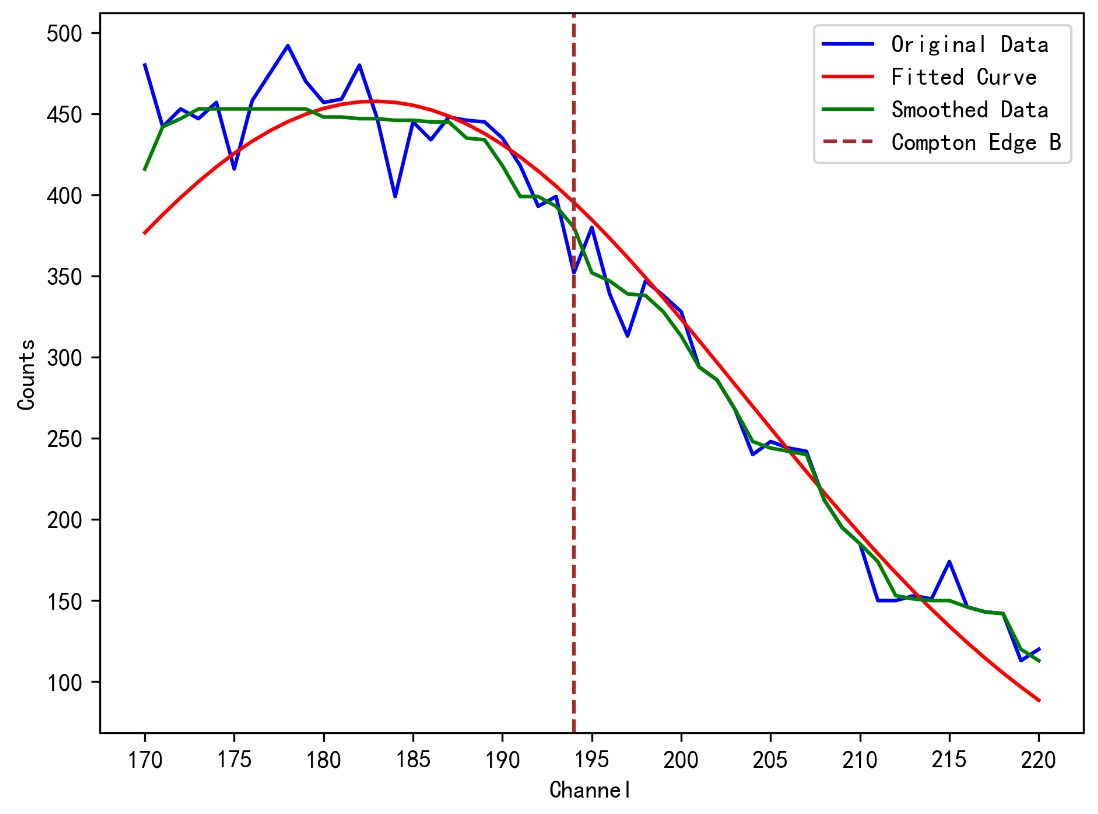}
        \caption{NaI and ${}^{137}$Cs Fitting}
    \end{subfigure}
    \caption{Compton Edge Fitting}
    \label{fig:fit2}
\end{figure}
The development of accurate and universally applicable calibration methods for scintillation detectors holds significant implications for various fields, including radiation dosimetry, environmental monitoring, and medical imaging. The proposed convolution model represents a significant advancement in this regard, offering a practical solution for calibrating scintillation detectors with predominantly Compton spectra.

The success of the convolution model paves the way for future research aimed at further refining and optimizing calibration techniques for scintillation detectors. Areas for potential improvement include exploring strategies for enhancing the robustness and adaptability of the model to accommodate various detector geometries, environmental conditions, and source characteristics.

Additionally, efforts to develop user-friendly software tools or standardized protocols for applying the convolution model could facilitate its widespread adoption and integration into existing calibration practices. By fostering collaboration and knowledge sharing within the scientific community, we can accelerate advancements in radiation detection technology and contribute to the improvement of public health and safety.

In conclusion, the proposed convolution model offers a promising solution to the challenges associated with calibrating scintillation detectors with predominantly Compton spectra. By providing accurate and universally applicable calibration methods, this model has the potential to enhance the precision and reliability of radiation detection and measurement across diverse applications, ultimately benefiting society as a whole.

In fitting to the actual function image, our fitting parameters are the Compton edge values B, the Gaussian function amplitude A, and the standard deviation of the Gaussian function $C = \sigma$.

According to the comparison between the theoretical curve and the experimental curve in \autoref{fig:fit1}, it can be seen that there are differences between the two in two energy regions:
1) Around $0.184 \mathrm{MeV}$, there is a backscattering peak in the experimental spectrum. This is due to the photoelectric effect of the scattered $\gamma$ photons returning to the flash peptone .
2) Within $0.430 \sim 0.480 \mathrm{MeV}$, the theoretical curve shows a faster upward trend than the experimental curve.

\begin{table}
    \centering
    \caption{BC408 Energy-Channel Calibration}
    \begin{tabular}{p{2cm}p{2.4cm}p{3.2cm}@{\hskip 0.5cm}p{2.6cm}}
    \hline\hline
         Radiation Source&  Gamma-ray Energy&  Maximum Energy of Recoil Electrons& Compton Edge Fitted Channel
\\ \hline
         Cs137&  662kev&  476.7kev& 2295
\\
         Co60&  1170kev&  963.2kev& 4670
\\
         Na22&  1275kev&  1061.2kev& 5030
\\ \hline \hline
    \end{tabular}
    \label{tab:tab1}
\end{table}

\begin{figure}[htbp]
    \centering
    \includegraphics[width=0.6\textwidth]{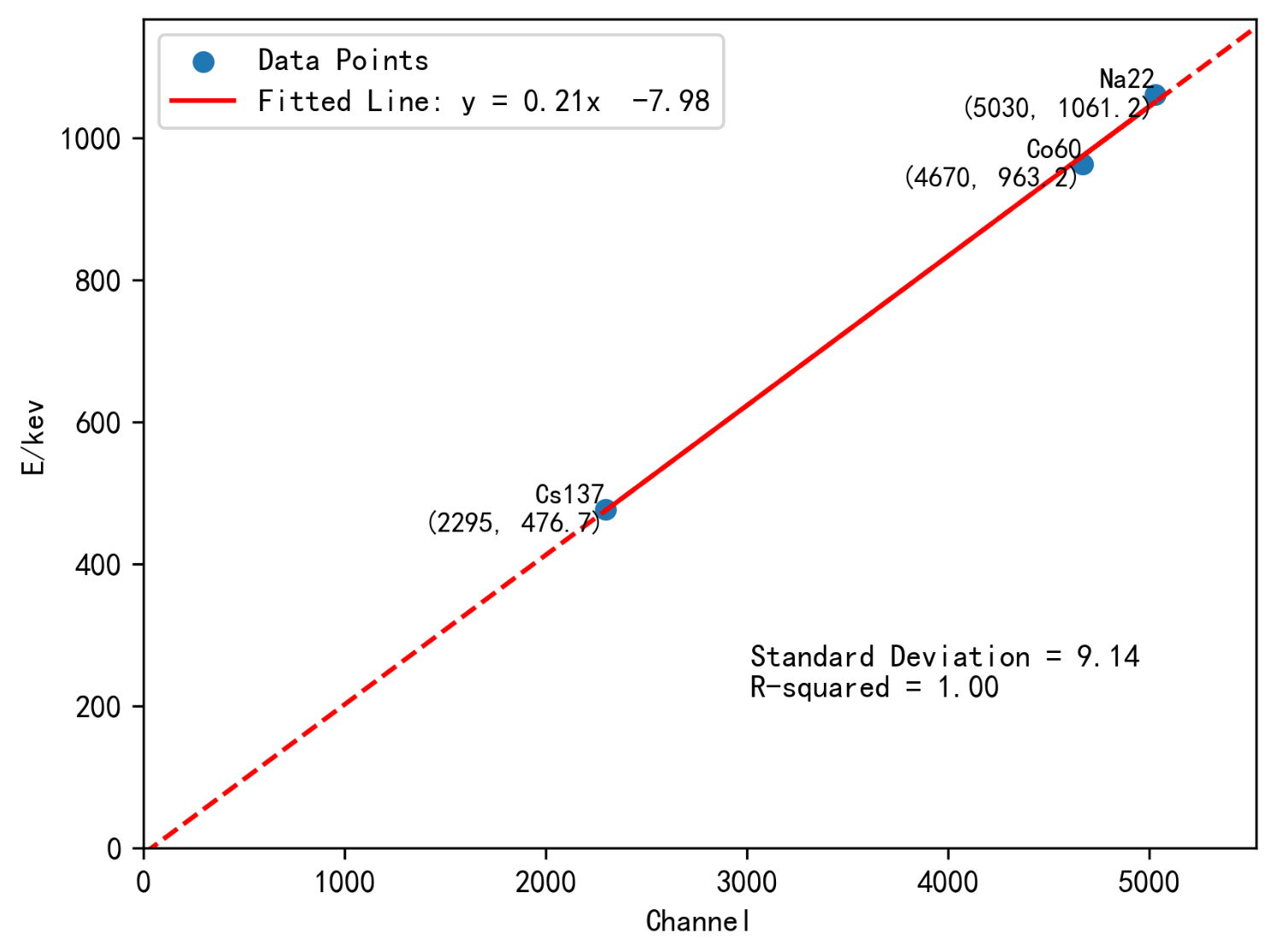}
    \caption{Compton Edge Fitting Verification}
    \label{fig:calib1}
\end{figure}

In order to study the difference between the theoretical curve and the experimental curve in this interval, the MCNP program is used to simulate the experimental process. The model is shown in \autoref{fig:fit2}. The simulation is done in two ways. When only the energy of $\gamma$ photons in the crystal is recorded During deposition, an unbroadened simulated spectrum is obtained. As can be seen from \autoref{fig:calib1}, without adding a broadening coefficient, the simulated energy spectrum is between the theoretical and experimental curves. This is because, during theoretical calculations, $\gamma$ is approximately considered  Photons scatter once in the crystal and leave the crystal, without considering the secondary effects of scattered photons. Due to the occurrence of secondary effects (including the photoelectric effect and secondary Compton scattering), some of the scattered photons are absorbed by the flash peptone. more remains in the crystal

\begin{figure}[htbp]
    \centering
    \begin{subfigure}{0.48\textwidth}
        \includegraphics[width=\textwidth]{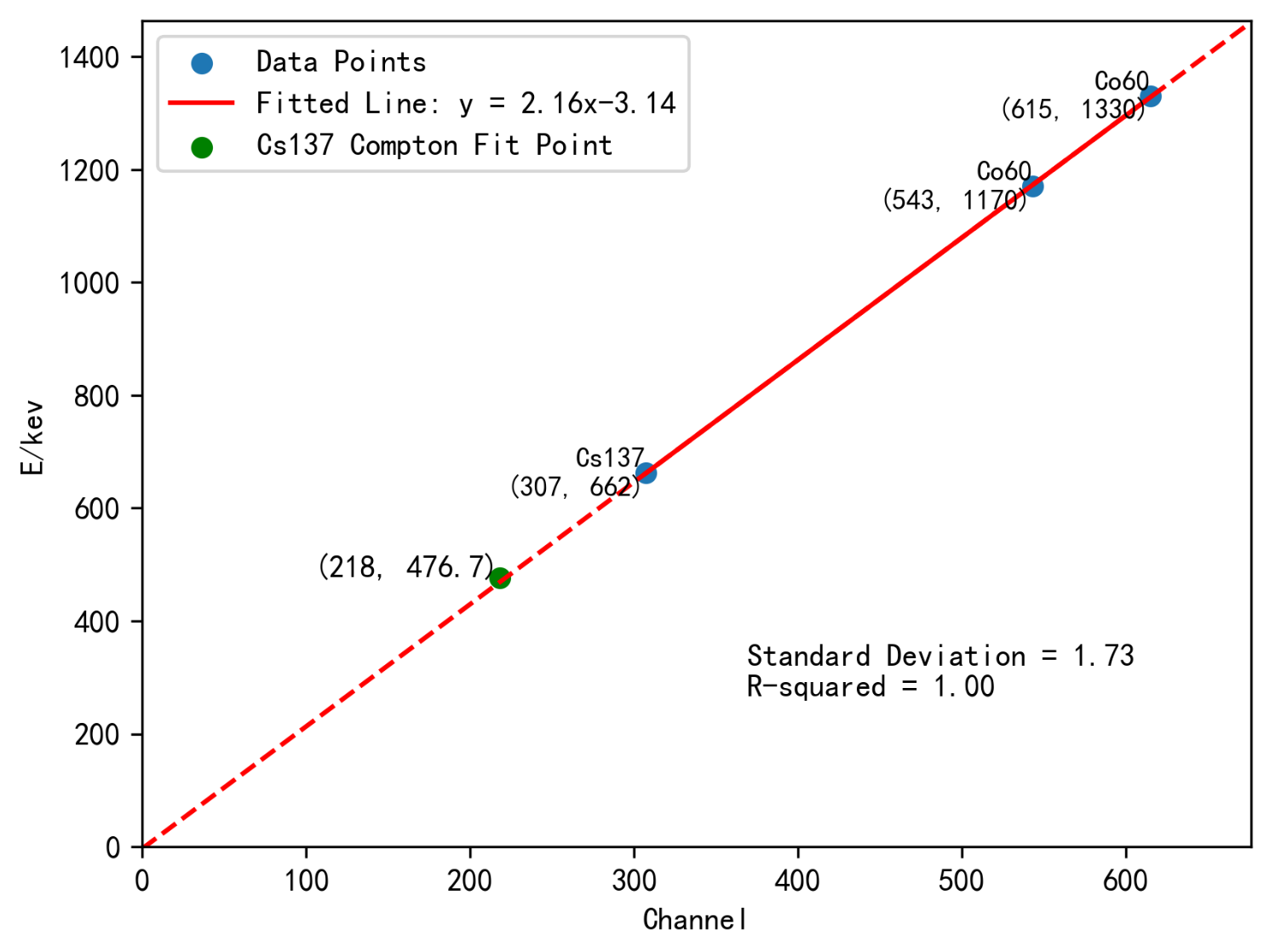}
        \caption{LaBr3}
    \end{subfigure}
    \begin{subfigure}{0.48\textwidth}
        \includegraphics[width=\textwidth]{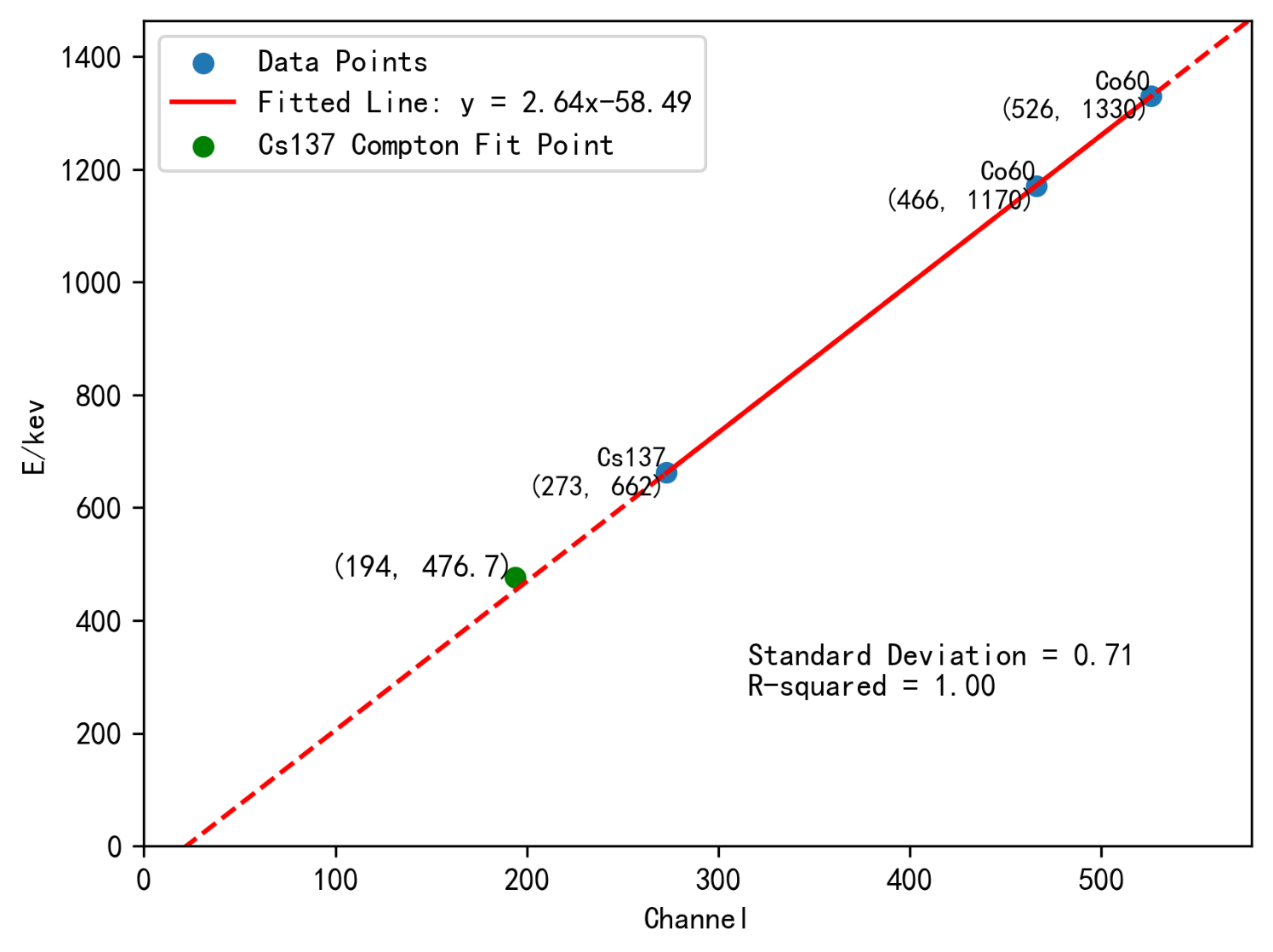}
        \caption{NaI}
    \end{subfigure}
    \caption{Compton Edge Fitting Verification}
    \label{fig:calib2}
\end{figure}

The simulation model of MCNP indicates that high energy causes the experimental curve to be lower than the theoretical curve. No changes in content have been made to the original text. However, the simulation curve without a broadening coefficient is still higher than the experimental curve. It is important to note that the text is free from grammatical errors, spelling mistakes, and punctuation errors, the $\mathrm{MCNP4C}$ processing card is used to calculate the pulse count in the program. The energy distribution is broadened by a Gaussian function, and the formula for calculating the half-maximum width of the energy peak $E_{\mathrm{FWHM}}$ is:
\begin{equation}
    E_{\mathrm{FWHM}}=a+b \sqrt{E_\gamma+c E_\gamma^2},
\end{equation}
Where $E_\gamma$ is the energy of $\gamma$ rays emitted by humans $(\mathrm{MeV})$. Extract ${ }^{152} \mathrm{Eu}$ and ${ }^{137} \mathrm{Cs}$ The energy half-peak width $E_{\mathrm{FWHM}}$ of the full-energy peak of the ${ }^{137} \mathrm{Cs}$ energy spectrum was fitted, and the coefficients $a=-6.8652, b=2.10147$ and $c=0.000123$ were obtained. As can be seen from \autoref{fig:calib2}, The broadened simulation curve is basically consistent with the experimental curve.

Ultimately, we compared the method via Compton calibration with the method at the top of Almighty Peak White Oh, and the final result proved to be an almost exact match, proving that our method is reliable, as shown in \autoref{fig:calib1} and \autoref{fig:calib2}. 

Therefore, the difference between the theoretical and experimental curves depends on factors such as the amplification of the photomultiplier tube and the statistical fluctuation of the number of flash peptone photons. Additionally, the peak-to-contemporary ratio of the simulation results is smaller than that of the experiment. This difference in the peak-to-combination ratio of the curve is due to a higher estimate of the laboratory background.

In summary, this paper uses the $\mathrm{NaI}(\mathrm{Tl})$ detector to measure the ${ }^{137} \mathrm{Cs}$ point source, and uses the Klein-Nishina Compton scattering differential cross-section formula to study The Compton plateau of the monoenergetic $\gamma$ ray energy spectrum was obtained. The results show that there is a big difference between the theoretical and experimental curves within $0.430 \sim 0.480 \mathrm{MeV}$. Through simulation calculations with the MCNP program, the photon in The energy spectrum broadening caused by factors such as the secondary effects in the flashing crystal and the amplification of the photomultiplier tube and the statistical fluctuation of the number of flashing photons resulted in simulation results that are in good agreement with the experimental energy spectrum.

\section{Discussion}

The proposed convolution model for fitting the energy-channel relationship of the Compton edge in scintillation detectors represents a significant advancement in calibration methodology. This discussion will delve into the implications, strengths, and limitations of the proposed model, as well as its broader significance in the field of radiation detection and measurement \cite{LOWE2000247}.

Firstly, the need for accurate calibration of scintillation detectors cannot be overstated. Calibration forms the foundation for precise energy measurement, which is crucial in various applications such as radiation dosimetry, environmental monitoring, and medical imaging. Traditionally, calibration methods have heavily relied on full-energy peaks, but as highlighted in the abstract, certain detectors, particularly plastic scintillation detectors, may not exhibit prominent full-energy peaks in their pulse height spectra. This limitation necessitates alternative calibration approaches, making the development of robust methods for calibrating Compton edges imperative.

The proposed convolution model addresses this challenge by providing a systematic and adaptable framework for calibrating the energy-channel relationship of the Compton edge across different types of detectors. By leveraging the unique characteristics of Compton plateaus and edges, the model offers a practical solution that circumvents the dependence on full-energy peaks, thereby widening the applicability of calibration techniques to a broader range of detectors. Furthermore, the experimental validation using various scintillator materials, including plastic scintillator BC408, NaI crystal, and LaBr$_3$ crystal, underscores the versatility and universality of the proposed model.

One of the notable strengths of the convolution model is its ability to achieve accurate calibration results with minimal manual intervention. Unlike traditional methods that may require subjective adjustments or manual selection of calibration points, the convolution model automates the calibration process, reducing the potential for human error and enhancing reproducibility. This automation not only streamlines the calibration procedure but also increases its reliability and robustness, particularly in scenarios where consistency and accuracy are paramount.

Moreover, the comparison of energy-channel functions obtained from the convolution model with those derived from full-energy peaks, revealing errors within 1\%, attests to the efficacy and precision of the proposed approach. Such close agreement between the calibration results validates the accuracy of the convolution model and instills confidence in its utility for practical applications \cite{GARDNER200487}.
However, it's important to acknowledge certain limitations and areas for future improvement. While the convolution model demonstrates promising performance across different detector types, its effectiveness may still be influenced by various factors such as detector geometry, environmental conditions, and source characteristics. Further research could explore strategies for enhancing the robustness and adaptability of the model to accommodate these variables more effectively.

Additionally, the practical implementation of the convolution model may require computational resources and expertise in signal processing techniques, which could pose challenges for some users. Efforts to develop user-friendly software tools or standardized protocols for applying the convolution model could facilitate its widespread adoption and integration into existing calibration practices.

One notable limitation of our study is the presence of significant background noise in the pulse height spectra, particularly in plastic scintillation detectors. This noise can obscure the Compton edge and affect the accuracy of calibration using the proposed convolution model. To address this limitation, future experiments could focus on mitigating background noise through improved shielding or signal processing techniques. Additionally, incorporating effective noise reduction methods into the convolution model may enhance its performance in noisy environments.

Furthermore, the sensitivity and robustness of the convolution model warrant further investigation. While our experiments demonstrate promising results in calibrating the energy-channel relationship of the Compton edge, additional testing under varying experimental conditions and with different detector configurations is necessary to assess the model's performance comprehensively. Future studies could explore sensitivity analyses to evaluate the model's response to variations in parameters such as detector geometry, energy resolution, and source characteristics.

Expanding the experimental dataset by conducting more groups of experiments would strengthen the validation and generalizability of our findings. Testing the convolution model with a wider range of scintillation detectors and radioactive sources can provide deeper insights into its applicability and universality. Moreover, increasing the sample size would allow for a more robust statistical analysis, enabling us to draw more reliable conclusions about the model's accuracy and performance across diverse experimental setups.

In addition to expanding the experimental data, future research directions could focus on refining the convolution model and exploring alternative calibration methods. Investigating advanced signal processing techniques or machine learning algorithms may offer opportunities to improve the accuracy and efficiency of energy-channel calibration in scintillation detectors. Collaborative efforts between researchers and industry stakeholders could facilitate the development and implementation of innovative calibration approaches that address the evolving needs of radiation detection applications.

In conclusion, our study presents a promising convolution model for accurately fitting the energy-channel relationship of the Compton edge in scintillation detectors. While our experiments verify the effectiveness of the model across different detector types, there are still areas for improvement and further investigation. By addressing the identified limitations, testing the model's sensitivity and robustness, and expanding the experimental dataset, we can enhance the reliability and applicability of the convolution model for calibration purposes. Continued research in this direction holds great potential for advancing the field of radiation detection and measurement, ultimately contributing to improved accuracy and reliability in various real-world applications.

\section{ACKNOWLEDGMENTS}

The authors gratefully acknowledge Qianru Zhao and Xiaoxu Chen for their insightful talks and input throughout the study.
This work is also supported by Sichuan University undergraduate students' platform for innovation and entrepreneurship training program (Grant No. C2024128729). 
The conclusions and analyses presented in this publication were produced using the following software: Python (Guido van Rossum 1986) \cite{10.5555/1593511}, Opencv (Intel, Willow Garage, Itseez) \cite{itseez2015opencv}, Scipy (Jones et al. 2001) \cite{2020SciPy-NMeth}, PyTorch  (Meta AI September 2016) \cite{NEURIPS2019_9015}, Matplotlib (Hunter 2007) \cite{Hunter:2007}, Seaborn \cite{Waskom2021} .
This work is finished on the server from Kun-Lun in Center for Theoretical Physics, School of Physics, Sichuan University.

\nocite{*}
\bibliography{ComptonEdge}

\begin{thebibliography}{20}%
\makeatletter
\providecommand \@ifxundefined [1]{%
 \@ifx{#1\undefined}
}%
\providecommand \@ifnum [1]{%
 \ifnum #1\expandafter \@firstoftwo
 \else \expandafter \@secondoftwo
 \fi
}%
\providecommand \@ifx [1]{%
 \ifx #1\expandafter \@firstoftwo
 \else \expandafter \@secondoftwo
 \fi
}%
\providecommand \natexlab [1]{#1}%
\providecommand \enquote  [1]{``#1''}%
\providecommand \bibnamefont  [1]{#1}%
\providecommand \bibfnamefont [1]{#1}%
\providecommand \citenamefont [1]{#1}%
\providecommand \href@noop [0]{\@secondoftwo}%
\providecommand \href [0]{\begingroup \@sanitize@url \@href}%
\providecommand \@href[1]{\@@startlink{#1}\@@href}%
\providecommand \@@href[1]{\endgroup#1\@@endlink}%
\providecommand \@sanitize@url [0]{\catcode `\\12\catcode `\$12\catcode `\&12\catcode `\#12\catcode `\^12\catcode `\_12\catcode `\%12\relax}%
\providecommand \@@startlink[1]{}%
\providecommand \@@endlink[0]{}%
\providecommand \url  [0]{\begingroup\@sanitize@url \@url }%
\providecommand \@url [1]{\endgroup\@href {#1}{\urlprefix }}%
\providecommand \urlprefix  [0]{URL }%
\providecommand \Eprint [0]{\href }%
\providecommand \doibase [0]{https://doi.org/}%
\providecommand \selectlanguage [0]{\@gobble}%
\providecommand \bibinfo  [0]{\@secondoftwo}%
\providecommand \bibfield  [0]{\@secondoftwo}%
\providecommand \translation [1]{[#1]}%
\providecommand \BibitemOpen [0]{}%
\providecommand \bibitemStop [0]{}%
\providecommand \bibitemNoStop [0]{.\EOS\space}%
\providecommand \EOS [0]{\spacefactor3000\relax}%
\providecommand \BibitemShut  [1]{\csname bibitem#1\endcsname}%
\let\auto@bib@innerbib\@empty
\bibitem [{\citenamefont {Wang}\ \emph {et~al.}(2023)\citenamefont {Wang}, \citenamefont {Dujardin}, \citenamefont {Freeman}, \citenamefont {Gehring}, \citenamefont {Hunter}, \citenamefont {Lecoq}, \citenamefont {Liu}, \citenamefont {Melcher}, \citenamefont {Morris}, \citenamefont {Nikl}, \citenamefont {Pilania}, \citenamefont {Pokharel}, \citenamefont {Robertson}, \citenamefont {Rutstrom}, \citenamefont {Sjue}, \citenamefont {Tremsin}, \citenamefont {Watson}, \citenamefont {Wiggins}, \citenamefont {Winch},\ and\ \citenamefont {Zhuravleva}}]{Wang_2023}%
  \BibitemOpen
  \bibfield  {author} {\bibinfo {author} {\bibfnamefont {Z.}~\bibnamefont {Wang}}, \bibinfo {author} {\bibfnamefont {C.}~\bibnamefont {Dujardin}}, \bibinfo {author} {\bibfnamefont {M.~S.}\ \bibnamefont {Freeman}}, \bibinfo {author} {\bibfnamefont {A.~E.}\ \bibnamefont {Gehring}}, \bibinfo {author} {\bibfnamefont {J.~F.}\ \bibnamefont {Hunter}}, \bibinfo {author} {\bibfnamefont {P.}~\bibnamefont {Lecoq}}, \bibinfo {author} {\bibfnamefont {W.}~\bibnamefont {Liu}}, \bibinfo {author} {\bibfnamefont {C.~L.}\ \bibnamefont {Melcher}}, \bibinfo {author} {\bibfnamefont {C.~L.}\ \bibnamefont {Morris}}, \bibinfo {author} {\bibfnamefont {M.}~\bibnamefont {Nikl}}, \bibinfo {author} {\bibfnamefont {G.}~\bibnamefont {Pilania}}, \bibinfo {author} {\bibfnamefont {R.}~\bibnamefont {Pokharel}}, \bibinfo {author} {\bibfnamefont {D.~G.}\ \bibnamefont {Robertson}}, \bibinfo {author} {\bibfnamefont {D.~J.}\ \bibnamefont {Rutstrom}}, \bibinfo {author} {\bibfnamefont {S.~K.}\ \bibnamefont {Sjue}}, \bibinfo {author} {\bibfnamefont {A.~S.}\ \bibnamefont {Tremsin}}, \bibinfo {author} {\bibfnamefont {S.~A.}\ \bibnamefont {Watson}}, \bibinfo {author} {\bibfnamefont {B.~W.}\ \bibnamefont {Wiggins}}, \bibinfo {author} {\bibfnamefont {N.~M.}\ \bibnamefont {Winch}},\ and\ \bibinfo {author} {\bibfnamefont {M.}~\bibnamefont {Zhuravleva}},\ }\bibfield  {title} {\bibinfo {title} {Needs, trends, and advances in scintillators for radiographic imaging and tomography},\ }\href {https://doi.org/10.1109/tns.2023.3290826} {\bibfield  {journal} {\bibinfo  {journal} {IEEE Transactions on Nuclear Science}\ }\textbf {\bibinfo {volume} {70}},\ \bibinfo {pages} {1244–1280} (\bibinfo {year} {2023})}\BibitemShut {NoStop}%
\bibitem [{\citenamefont {Compton}(1923{\natexlab{a}})}]{PhysRev.21.483}%
  \BibitemOpen
  \bibfield  {author} {\bibinfo {author} {\bibfnamefont {A.~H.}\ \bibnamefont {Compton}},\ }\bibfield  {title} {\bibinfo {title} {A quantum theory of the scattering of x-rays by light elements},\ }\href {https://doi.org/10.1103/PhysRev.21.483} {\bibfield  {journal} {\bibinfo  {journal} {Phys. Rev.}\ }\textbf {\bibinfo {volume} {21}},\ \bibinfo {pages} {483} (\bibinfo {year} {1923}{\natexlab{a}})}\BibitemShut {NoStop}%
\bibitem [{\citenamefont {Compton}(1923{\natexlab{b}})}]{PhysRev.22.409}%
  \BibitemOpen
  \bibfield  {author} {\bibinfo {author} {\bibfnamefont {A.~H.}\ \bibnamefont {Compton}},\ }\bibfield  {title} {\bibinfo {title} {The spectrum of scattered x-rays},\ }\href {https://doi.org/10.1103/PhysRev.22.409} {\bibfield  {journal} {\bibinfo  {journal} {Phys. Rev.}\ }\textbf {\bibinfo {volume} {22}},\ \bibinfo {pages} {409} (\bibinfo {year} {1923}{\natexlab{b}})}\BibitemShut {NoStop}%
\bibitem [{\citenamefont {Siciliano}\ \emph {et~al.}(2008)\citenamefont {Siciliano}, \citenamefont {Ely}, \citenamefont {Kouzes}, \citenamefont {Schweppe}, \citenamefont {Strachan},\ and\ \citenamefont {Yokuda}}]{article2}%
  \BibitemOpen
  \bibfield  {author} {\bibinfo {author} {\bibfnamefont {E.}~\bibnamefont {Siciliano}}, \bibinfo {author} {\bibfnamefont {J.}~\bibnamefont {Ely}}, \bibinfo {author} {\bibfnamefont {R.}~\bibnamefont {Kouzes}}, \bibinfo {author} {\bibfnamefont {J.}~\bibnamefont {Schweppe}}, \bibinfo {author} {\bibfnamefont {D.}~\bibnamefont {Strachan}},\ and\ \bibinfo {author} {\bibfnamefont {S.}~\bibnamefont {Yokuda}},\ }\bibfield  {title} {\bibinfo {title} {Energy calibration of gamma spectra in plastic scintillators using compton kinematics},\ }\href {https://doi.org/10.1016/j.nima.2008.06.031} {\bibfield  {journal} {\bibinfo  {journal} {Nuclear Instruments and Methods in Physics Research Section A: Accelerators, Spectrometers, Detectors and Associated Equipment}\ }\textbf {\bibinfo {volume} {594}},\ \bibinfo {pages} {232–243} (\bibinfo {year} {2008})}\BibitemShut {NoStop}%
\bibitem [{\citenamefont {Deng}\ \emph {et~al.}(2022)\citenamefont {Deng}, \citenamefont {Zhang},\ and\ \citenamefont {Sun}}]{Deng_2022}%
  \BibitemOpen
  \bibfield  {author} {\bibinfo {author} {\bibfnamefont {J.-G.}\ \bibnamefont {Deng}}, \bibinfo {author} {\bibfnamefont {H.-F.}\ \bibnamefont {Zhang}},\ and\ \bibinfo {author} {\bibfnamefont {X.-D.}\ \bibnamefont {Sun}},\ }\bibfield  {title} {\bibinfo {title} {New behaviors of $\alpha$-particle preformation factors near doubly magic 100sn *},\ }\href {https://doi.org/10.1088/1674-1137/ac5a9f} {\bibfield  {journal} {\bibinfo  {journal} {Chinese Physics C}\ }\textbf {\bibinfo {volume} {46}},\ \bibinfo {pages} {061001} (\bibinfo {year} {2022})}\BibitemShut {NoStop}%
\bibitem [{\citenamefont {Jeon}\ \emph {et~al.}(2019)\citenamefont {Jeon}, \citenamefont {Kim}, \citenamefont {Moon},\ and\ \citenamefont {Cho}}]{JEON20198}%
  \BibitemOpen
  \bibfield  {author} {\bibinfo {author} {\bibfnamefont {B.}~\bibnamefont {Jeon}}, \bibinfo {author} {\bibfnamefont {J.}~\bibnamefont {Kim}}, \bibinfo {author} {\bibfnamefont {M.}~\bibnamefont {Moon}},\ and\ \bibinfo {author} {\bibfnamefont {G.}~\bibnamefont {Cho}},\ }\bibfield  {title} {\bibinfo {title} {Parametric optimization for energy calibration and gamma response function of plastic scintillation detectors using a genetic algorithm},\ }\href {https://doi.org/https://doi.org/10.1016/j.nima.2019.03.003} {\bibfield  {journal} {\bibinfo  {journal} {Nuclear Instruments and Methods in Physics Research Section A: Accelerators, Spectrometers, Detectors and Associated Equipment}\ }\textbf {\bibinfo {volume} {930}},\ \bibinfo {pages} {8} (\bibinfo {year} {2019})}\BibitemShut {NoStop}%
\bibitem [{\citenamefont {Jeon}\ \emph {et~al.}(2023)\citenamefont {Jeon}, \citenamefont {Park}, \citenamefont {Park}, \citenamefont {Lee},\ and\ \citenamefont {Moon}}]{article3}%
  \BibitemOpen
  \bibfield  {author} {\bibinfo {author} {\bibfnamefont {B.}~\bibnamefont {Jeon}}, \bibinfo {author} {\bibfnamefont {J.}~\bibnamefont {Park}}, \bibinfo {author} {\bibfnamefont {K.}~\bibnamefont {Park}}, \bibinfo {author} {\bibfnamefont {J.}~\bibnamefont {Lee}},\ and\ \bibinfo {author} {\bibfnamefont {M.}~\bibnamefont {Moon}},\ }\bibfield  {title} {\bibinfo {title} {Automatic energy calibration of radiation portal monitors using deep learning and spectral remapping},\ }\href {https://doi.org/10.1088/1748-0221/18/01/P01031} {\bibfield  {journal} {\bibinfo  {journal} {Journal of Instrumentation}\ }\textbf {\bibinfo {volume} {18}},\ \bibinfo {pages} {P01031}}\BibitemShut {NoStop}%
\bibitem [{\citenamefont {Abbas}(2006)}]{Abbas_2006}%
  \BibitemOpen
  \bibfield  {author} {\bibinfo {author} {\bibfnamefont {M.~I.}\ \bibnamefont {Abbas}},\ }\bibfield  {title} {\bibinfo {title} {Hpge detector absolute full-energy peak efficiency calibration including coincidence correction for circular disc sources},\ }\href {https://doi.org/10.1088/0022-3727/39/18/005} {\bibfield  {journal} {\bibinfo  {journal} {Journal of Physics D: Applied Physics}\ }\textbf {\bibinfo {volume} {39}},\ \bibinfo {pages} {3952} (\bibinfo {year} {2006})}\BibitemShut {NoStop}%
\bibitem [{\citenamefont {Stokes}(2002)}]{article}%
  \BibitemOpen
  \bibfield  {author} {\bibinfo {author} {\bibfnamefont {A.}~\bibnamefont {Stokes}},\ }\bibfield  {title} {\bibinfo {title} {Numerical fourier-analysis method for correction of widths and shapes of lines on x-ray powder photographs},\ }\href {https://doi.org/10.1088/0959-5309/61/4/311} {\bibfield  {journal} {\bibinfo  {journal} {Proceedings of the Physical Society}\ }\textbf {\bibinfo {volume} {61}},\ \bibinfo {pages} {382} (\bibinfo {year} {2002})}\BibitemShut {NoStop}%
\bibitem [{\citenamefont {Paatero}\ \emph {et~al.}(1974)\citenamefont {Paatero}, \citenamefont {Manninen},\ and\ \citenamefont {Paakkari}}]{Paatero1974-PAADIC}%
  \BibitemOpen
  \bibfield  {author} {\bibinfo {author} {\bibfnamefont {P.}~\bibnamefont {Paatero}}, \bibinfo {author} {\bibfnamefont {S.}~\bibnamefont {Manninen}},\ and\ \bibinfo {author} {\bibfnamefont {T.}~\bibnamefont {Paakkari}},\ }\bibfield  {title} {\bibinfo {title} {Deconvolution in compton profile measurements},\ }\href {https://doi.org/10.1080/14786437408207281} {\bibfield  {journal} {\bibinfo  {journal} {Philosophical Magazine}\ }\textbf {\bibinfo {volume} {30}},\ \bibinfo {pages} {1281} (\bibinfo {year} {1974})}\BibitemShut {NoStop}%
\bibitem [{\citenamefont {Philip}(1983)}]{Philip1983DeconvolutionOG}%
  \BibitemOpen
  \bibfield  {author} {\bibinfo {author} {\bibfnamefont {J.}~\bibnamefont {Philip}},\ }\bibfield  {title} {\bibinfo {title} {Deconvolution of gaussian and other kernels}\ }(\bibinfo {year} {1983})\BibitemShut {NoStop}%
\bibitem [{\citenamefont {Lowe}(2000)}]{LOWE2000247}%
  \BibitemOpen
  \bibfield  {author} {\bibinfo {author} {\bibfnamefont {B.}~\bibnamefont {Lowe}},\ }\bibfield  {title} {\bibinfo {title} {An analytical description of low-energy x-ray spectra in si(li) and hpge detectors},\ }\href {https://doi.org/https://doi.org/10.1016/S0168-9002(99)00933-X} {\bibfield  {journal} {\bibinfo  {journal} {Nuclear Instruments and Methods in Physics Research Section A: Accelerators, Spectrometers, Detectors and Associated Equipment}\ }\textbf {\bibinfo {volume} {439}},\ \bibinfo {pages} {247} (\bibinfo {year} {2000})}\BibitemShut {NoStop}%
\bibitem [{\citenamefont {Gardner}\ and\ \citenamefont {Sood}(2004)}]{GARDNER200487}%
  \BibitemOpen
  \bibfield  {author} {\bibinfo {author} {\bibfnamefont {R.~P.}\ \bibnamefont {Gardner}}\ and\ \bibinfo {author} {\bibfnamefont {A.}~\bibnamefont {Sood}},\ }\bibfield  {title} {\bibinfo {title} {A monte carlo simulation approach for generating nai detector response functions (drfs) that accounts for non-linearity and variable flat continua},\ }\href {https://doi.org/https://doi.org/10.1016/S0168-583X(03)01539-8} {\bibfield  {journal} {\bibinfo  {journal} {Nuclear Instruments and Methods in Physics Research Section B: Beam Interactions with Materials and Atoms}\ }\textbf {\bibinfo {volume} {213}},\ \bibinfo {pages} {87} (\bibinfo {year} {2004})},\ \bibinfo {note} {5th Topical Meeting on Industrial Radiation and Radioisotope Measurement Applications}\BibitemShut {NoStop}%
\bibitem [{\citenamefont {Van~Rossum}\ and\ \citenamefont {Drake}(2009)}]{10.5555/1593511}%
  \BibitemOpen
  \bibfield  {author} {\bibinfo {author} {\bibfnamefont {G.}~\bibnamefont {Van~Rossum}}\ and\ \bibinfo {author} {\bibfnamefont {F.~L.}\ \bibnamefont {Drake}},\ }\href@noop {} {\emph {\bibinfo {title} {Python 3 Reference Manual}}}\ (\bibinfo  {publisher} {CreateSpace},\ \bibinfo {address} {Scotts Valley, CA},\ \bibinfo {year} {2009})\BibitemShut {NoStop}%
\bibitem [{\citenamefont {Itseez}(2015)}]{itseez2015opencv}%
  \BibitemOpen
  \bibfield  {author} {\bibinfo {author} {\bibnamefont {Itseez}},\ }\href@noop {} {\bibinfo {title} {Open source computer vision library}},\ \bibinfo {howpublished} {\url{https://github.com/itseez/opencv}} (\bibinfo {year} {2015})\BibitemShut {NoStop}%
\bibitem [{\citenamefont {Virtanen}\ \emph {et~al.}(2020)\citenamefont {Virtanen}, \citenamefont {Gommers}, \citenamefont {Oliphant}, \citenamefont {Haberland}, \citenamefont {Reddy}, \citenamefont {Cournapeau}, \citenamefont {Burovski}, \citenamefont {Peterson}, \citenamefont {Weckesser}, \citenamefont {Bright}, \citenamefont {{van der Walt}}, \citenamefont {Brett}, \citenamefont {Wilson}, \citenamefont {Millman}, \citenamefont {Mayorov}, \citenamefont {Nelson}, \citenamefont {Jones}, \citenamefont {Kern}, \citenamefont {Larson}, \citenamefont {Carey}, \citenamefont {Polat}, \citenamefont {Feng}, \citenamefont {Moore}, \citenamefont {{VanderPlas}}, \citenamefont {Laxalde}, \citenamefont {Perktold}, \citenamefont {Cimrman}, \citenamefont {Henriksen}, \citenamefont {Quintero}, \citenamefont {Harris}, \citenamefont {Archibald}, \citenamefont {Ribeiro}, \citenamefont {Pedregosa}, \citenamefont {{van Mulbregt}},\ and\ \citenamefont {{SciPy 1.0 Contributors}}}]{2020SciPy-NMeth}%
  \BibitemOpen
  \bibfield  {author} {\bibinfo {author} {\bibfnamefont {P.}~\bibnamefont {Virtanen}}, \bibinfo {author} {\bibfnamefont {R.}~\bibnamefont {Gommers}}, \bibinfo {author} {\bibfnamefont {T.~E.}\ \bibnamefont {Oliphant}}, \bibinfo {author} {\bibfnamefont {M.}~\bibnamefont {Haberland}}, \bibinfo {author} {\bibfnamefont {T.}~\bibnamefont {Reddy}}, \bibinfo {author} {\bibfnamefont {D.}~\bibnamefont {Cournapeau}}, \bibinfo {author} {\bibfnamefont {E.}~\bibnamefont {Burovski}}, \bibinfo {author} {\bibfnamefont {P.}~\bibnamefont {Peterson}}, \bibinfo {author} {\bibfnamefont {W.}~\bibnamefont {Weckesser}}, \bibinfo {author} {\bibfnamefont {J.}~\bibnamefont {Bright}}, \bibinfo {author} {\bibfnamefont {S.~J.}\ \bibnamefont {{van der Walt}}}, \bibinfo {author} {\bibfnamefont {M.}~\bibnamefont {Brett}}, \bibinfo {author} {\bibfnamefont {J.}~\bibnamefont {Wilson}}, \bibinfo {author} {\bibfnamefont {K.~J.}\ \bibnamefont {Millman}}, \bibinfo {author} {\bibfnamefont {N.}~\bibnamefont {Mayorov}}, \bibinfo {author} {\bibfnamefont {A.~R.~J.}\ \bibnamefont {Nelson}}, \bibinfo {author} {\bibfnamefont {E.}~\bibnamefont {Jones}}, \bibinfo {author} {\bibfnamefont {R.}~\bibnamefont {Kern}}, \bibinfo {author} {\bibfnamefont {E.}~\bibnamefont {Larson}}, \bibinfo {author} {\bibfnamefont {C.~J.}\ \bibnamefont {Carey}}, \bibinfo {author} {\bibfnamefont {{\.I}.}~\bibnamefont {Polat}}, \bibinfo {author} {\bibfnamefont {Y.}~\bibnamefont {Feng}}, \bibinfo {author} {\bibfnamefont {E.~W.}\ \bibnamefont {Moore}}, \bibinfo {author} {\bibfnamefont {J.}~\bibnamefont {{VanderPlas}}}, \bibinfo {author} {\bibfnamefont {D.}~\bibnamefont {Laxalde}}, \bibinfo {author} {\bibfnamefont {J.}~\bibnamefont {Perktold}}, \bibinfo {author} {\bibfnamefont {R.}~\bibnamefont {Cimrman}}, \bibinfo {author} {\bibfnamefont {I.}~\bibnamefont {Henriksen}}, \bibinfo {author} {\bibfnamefont {E.~A.}\ \bibnamefont {Quintero}}, \bibinfo {author} {\bibfnamefont {C.~R.}\ \bibnamefont {Harris}}, \bibinfo {author} {\bibfnamefont {A.~M.}\ \bibnamefont {Archibald}}, \bibinfo {author} {\bibfnamefont {A.~H.}\ \bibnamefont {Ribeiro}}, \bibinfo {author} {\bibfnamefont {F.}~\bibnamefont {Pedregosa}}, \bibinfo {author} {\bibfnamefont {P.}~\bibnamefont {{van Mulbregt}}},\ and\ \bibinfo {author} {\bibnamefont {{SciPy 1.0 Contributors}}},\ }\bibfield  {title} {\bibinfo {title} {{{SciPy} 1.0: Fundamental Algorithms for Scientific Computing in Python}},\ }\href {https://doi.org/10.1038/s41592-019-0686-2} {\bibfield  {journal} {\bibinfo  {journal} {Nature Methods}\ }\textbf {\bibinfo {volume} {17}},\ \bibinfo {pages} {261} (\bibinfo {year} {2020})}\BibitemShut {NoStop}%
\bibitem [{\citenamefont {Paszke}\ \emph {et~al.}(2019)\citenamefont {Paszke}, \citenamefont {Gross}, \citenamefont {Massa}, \citenamefont {Lerer}, \citenamefont {Bradbury}, \citenamefont {Chanan}, \citenamefont {Killeen}, \citenamefont {Lin}, \citenamefont {Gimelshein}, \citenamefont {Antiga}, \citenamefont {Desmaison}, \citenamefont {Kopf}, \citenamefont {Yang}, \citenamefont {DeVito}, \citenamefont {Raison}, \citenamefont {Tejani}, \citenamefont {Chilamkurthy}, \citenamefont {Steiner}, \citenamefont {Fang}, \citenamefont {Bai},\ and\ \citenamefont {Chintala}}]{NEURIPS2019_9015}%
  \BibitemOpen
  \bibfield  {author} {\bibinfo {author} {\bibfnamefont {A.}~\bibnamefont {Paszke}}, \bibinfo {author} {\bibfnamefont {S.}~\bibnamefont {Gross}}, \bibinfo {author} {\bibfnamefont {F.}~\bibnamefont {Massa}}, \bibinfo {author} {\bibfnamefont {A.}~\bibnamefont {Lerer}}, \bibinfo {author} {\bibfnamefont {J.}~\bibnamefont {Bradbury}}, \bibinfo {author} {\bibfnamefont {G.}~\bibnamefont {Chanan}}, \bibinfo {author} {\bibfnamefont {T.}~\bibnamefont {Killeen}}, \bibinfo {author} {\bibfnamefont {Z.}~\bibnamefont {Lin}}, \bibinfo {author} {\bibfnamefont {N.}~\bibnamefont {Gimelshein}}, \bibinfo {author} {\bibfnamefont {L.}~\bibnamefont {Antiga}}, \bibinfo {author} {\bibfnamefont {A.}~\bibnamefont {Desmaison}}, \bibinfo {author} {\bibfnamefont {A.}~\bibnamefont {Kopf}}, \bibinfo {author} {\bibfnamefont {E.}~\bibnamefont {Yang}}, \bibinfo {author} {\bibfnamefont {Z.}~\bibnamefont {DeVito}}, \bibinfo {author} {\bibfnamefont {M.}~\bibnamefont {Raison}}, \bibinfo {author} {\bibfnamefont {A.}~\bibnamefont {Tejani}}, \bibinfo {author} {\bibfnamefont {S.}~\bibnamefont {Chilamkurthy}}, \bibinfo {author} {\bibfnamefont {B.}~\bibnamefont {Steiner}}, \bibinfo {author} {\bibfnamefont {L.}~\bibnamefont {Fang}}, \bibinfo {author} {\bibfnamefont {J.}~\bibnamefont {Bai}},\ and\ \bibinfo {author} {\bibfnamefont {S.}~\bibnamefont {Chintala}},\ }\bibfield  {title} {\bibinfo {title} {Pytorch: An imperative style, high-performance deep learning library},\ }in\ \href {http://papers.neurips.cc/paper/9015-pytorch-an-imperative-style-high-performance-deep-learning-library.pdf} {\emph {\bibinfo {booktitle} {Advances in Neural Information Processing Systems 32}}}\ (\bibinfo  {publisher} {Curran Associates, Inc.},\ \bibinfo {year} {2019})\ pp.\ \bibinfo {pages} {8024--8035}\BibitemShut {NoStop}%
\bibitem [{\citenamefont {Hunter}(2007)}]{Hunter:2007}%
  \BibitemOpen
  \bibfield  {author} {\bibinfo {author} {\bibfnamefont {J.~D.}\ \bibnamefont {Hunter}},\ }\bibfield  {title} {\bibinfo {title} {Matplotlib: A 2d graphics environment},\ }\href {https://doi.org/10.1109/MCSE.2007.55} {\bibfield  {journal} {\bibinfo  {journal} {Computing in Science \& Engineering}\ }\textbf {\bibinfo {volume} {9}},\ \bibinfo {pages} {90} (\bibinfo {year} {2007})}\BibitemShut {NoStop}%
\bibitem [{\citenamefont {Waskom}(2021)}]{Waskom2021}%
  \BibitemOpen
  \bibfield  {author} {\bibinfo {author} {\bibfnamefont {M.~L.}\ \bibnamefont {Waskom}},\ }\bibfield  {title} {\bibinfo {title} {seaborn: statistical data visualization},\ }\href {https://doi.org/10.21105/joss.03021} {\bibfield  {journal} {\bibinfo  {journal} {Journal of Open Source Software}\ }\textbf {\bibinfo {volume} {6}},\ \bibinfo {pages} {3021} (\bibinfo {year} {2021})}\BibitemShut {NoStop}%
\bibitem [{\citenamefont {Qiao}\ \emph {et~al.}(2021)\citenamefont {Qiao}, \citenamefont {Wei},\ and\ \citenamefont {Chen}}]{cryst11050525}%
  \BibitemOpen
  \bibfield  {author} {\bibinfo {author} {\bibfnamefont {C.-K.}\ \bibnamefont {Qiao}}, \bibinfo {author} {\bibfnamefont {J.-W.}\ \bibnamefont {Wei}},\ and\ \bibinfo {author} {\bibfnamefont {L.}~\bibnamefont {Chen}},\ }\bibfield  {title} {\bibinfo {title} {An overview of the compton scattering calculation},\ }\bibfield  {journal} {\bibinfo  {journal} {Crystals}\ }\textbf {\bibinfo {volume} {11}},\ \href {https://doi.org/10.3390/cryst11050525} {10.3390/cryst11050525} (\bibinfo {year} {2021})\BibitemShut {NoStop}%
\end{thebibliography}%

\end{document}